\documentclass[journal=jctcee,manuscript=article]{achemso}
\setkeys{acs}{usetitle=true}
\usepackage[version=3]{mhchem} % Formula subscripts using \ce{}terials

\usepackage[version=3]{mhchem} % Formula subscripts using \ce{}

\usepackage{graphicx}% Include figure files
\usepackage{dcolumn}% Align table columns on decimal point
\usepackage{bm}% bold math
\usepackage{amsmath}%math
\usepackage{commath}%math formatting
\usepackage[dvipsnames]{xcolor}
\usepackage{placeins} % FloatBarrier
\usepackage{dblfloatfix}

\usepackage[utf8]{inputenc}
\usepackage[T1]{fontenc}
\usepackage{mathptmx}
\usepackage{etoolbox}
\usepackage{bbold}
\usepackage{hyperref}

\usepackage{ulem}

\title{Quantum Monte Carlo Benchmarking of Molecular Adsorption on Graphene-Supported Single Pt Atom} 

\author{Jeonghwan Ahn}
\affiliation{The Anthony J Leggett Institute for Condensed Matter Theory, Department of Physics, University of Illinois at Urbana-Champaign, Urbana, Illinois 61801, USA}
\alsoaffiliation{Department of Physics, Konkuk University, Seoul 05029, Korea}
\author{Iuegyun Hong}
\affiliation{Department of Physics, Konkuk University, Seoul 05029, Korea}
\author{Gwangyoung Lee}
\affiliation{Department of Physics, Konkuk University, Seoul 05029, Korea}
\author{Hyeondeok Shin}
\affiliation{Computational Science Division, Argonne National Laboratory, Argonne, Illinois 60439, USA}
\author{Anouar Benali}
\affiliation{Computational Science Division, Argonne National Laboratory, Argonne, Illinois 60439, USA}
\author{Yongkyung Kwon}
\email{ykwon@konkuk.ac.kr}
\affiliation{Department of Physics, Konkuk University, Seoul 05029, Korea}

\begin{document}

\begin{abstract}
{
The precise understanding of adsorption energetics and molecular geometry at catalytic sites is fundamental for advancing catalysis, particularly under the constraints of resource efficiency and environmental sustainability. This study benchmarks the performance of density functional theory (DFT) calculations against diffusion Monte Carlo (DMC) calculations for adsorption properties of small gas molecules relevant to CO oxidation---namely O$_2$, CO, CO$_2$, and atomic oxygen---on
%(O$_2$, CO, and CO$_2$) and atomic oxygen, which are involved in the CO oxidation process, on 
a single Pt atom supported by pristine graphene. 
Our findings reveal that DMC calculations provide a significantly different landscape of adsorption energetics compared to DFT results. Notably, DFT predicts different lowest-energy configurations and spin states, particularly for O$_2$, which suggests potential discrepancies in predicting the catalytic behavior. Furthermore, this study identifies the critical issue of CO poisoning, highlighted by the large disparity between the DMC adsorption energies of O$_2$ ($-1.23(2)$ eV) and CO ($-3.37(1)$ eV),
%by DMC adsorption energies of -1.23(2) eV for O$_2$ and -3.37(1) eV for CO, 
which can inhibit the catalytic process. 
These results emphasize the necessity for more sophisticated computational approaches in catalysis research, aiming to refine the prediction accuracy of reaction mechanisms and to enhance the design of more effective catalysts.
}
\end{abstract}

\maketitle

%\begin{multicols}{2}
%\narrowtext

\section{Introduction}
\label{sec:intro}
Catalysis plays an indispensable role in addressing the challenges of the current energy crisis, particularly through its applications in industrial processes. Platinum (Pt)-based catalysts are pivotal in this regard, serving critical functions in power generation and automotive technologies. Notably, they are utilized in catalytic converters to mitigate exhaust emissions and in fuel cells to enhance energy efficiency. Despite their widespread use, traditional Pt-based catalysts are hindered by the limited participation of Pt atoms in the catalytic reactions and by high production costs~\cite{zhang2019pt}. These limitations underscore the urgent need for innovative alternatives.
Graphene-supported Pt catalysts have emerged as a promising solution. They not only offer a larger effective reaction area compared to traditional catalysts, which reduces the required amount of Pt, but also allow for nanoscale design control of the catalyst structure~\cite{kong06,okazaki10,sun13,cheng16}.  
Even a single-atom catalysis has been explored to maximize the efficiency of the Pt catalyst~\cite{sun13}.

A fundamental understanding of adsorption energetics and molecular geometry at active sites is essential to optimizing catalyst design. These factors influence catalytic reaction pathways, thermodynamic properties, and electrochemical performance. 
For example, the mechanism of the CO oxidation may vary depending on the relative adsorption energies of CO and O$_2$ molecules. 
In the Eley–Rideal (ER) mechanism, CO molecules reacts directly with preadsorbed oxygen species without first adsorbing onto the substrate. This pathway is favored when the  adsorption energy of O$_2$ exceeds that of CO.
Conversely, the Langmuir–Hinshelwood(LH) mechanism involves the reaction of CO and O$_2$ molecules coadsorbed on neighboring sites, typically when the CO adsorption energy is comparable to or greater than that of O$_2$. On the other hand, 
strong CO adsorption could lead to a phenomenon known as CO poisoning, which severely impairs the catalyst's functionality by blocking active sites.
These insights highlight the importance of accurate calculations of adsorption energetics in understanding and improving catalytic mechanisms.

Most of previous molecular adsorption studies have primarily employed the nudged elastic band (NEB) method and density functional theory (DFT) calculations~\cite{tang12,lim12,liu14,tang15}.
However, these methods have limitations in capturing many-body correlation effects, which are influenced by factors such as system dimensionality~\cite{shin2014cohesion}, molecular geometry~\cite{shin2019O2,lee2022hydrogen}, and spin states~\cite{ahn21}.
Therefore, there is a pressing need for high-fidelity computational methods that can account for these effects to provide more reliable predictions of catalytic properties. 
To address these challenges, this study employs diffusion Monte Carlo (DMC) calculations, a robust many-body technique based on stochastic processes that solve the Schr\"{o}dinger equation directly~\cite{foulkes01}. DMC has demonstrated exceptional accuracy in determining cohesion and adsorption energetics for low-dimensional materials~\cite{mostaani15,shulenburger15,shin17,ahn18,ahn20,ahn21,ahn21b,ahn2023structural,ahn2024exploring,wines2024towards,ahn2025stacking}, offering new insights into the adsorption processes and enhancing our understanding of catalytic mechanisms at the atomic level.

This work is focused on the molecular adsorption on a single Pt atom supported by a graphene surface.
By considering spin degrees of freedom and multiple molecular geometries, we have computed the adsorption energies of  gas molecules (O$_{2}$, CO, CO$_{2}$, and atomic oxygen) which are involved in the CO oxidation process. 
Compared to the DMC results, the DFT-PBE calculations are found to significantly overestimate the molecular adsorption energies.
Notably, for the O$_2$ adsorption, the lowest-energy geometries and spin states predicted by DFT and DMC are not in agreement with each other, manifesting the significance of many-body correlation effects in determining the adsorption energetics. This suggests that many-body energy landscape would be different from the one used in the DFT-based NEB method, calling for a careful choice of initial and final configurations of NEB calculations to determine reaction paths and activation energies.

\section{Methodology}
\label{sec:method}

We conducted fixed-node DMC calculations using the QMCPACK code~\cite{kim18,kent20}. For trial wave functions, we used Slater-Jastrow types with Jastrow factors up to three-body correlations, namely, electron-electron-ion correlations. The Kohn-Sham orbitals constituting the Slater determinants were obtained from spin-polarized DFT calculations based on the PBE+$U$ functional~\cite{perdew96} as implemented in QUANTUM ESPRESSO package~\cite{giannozzi09}. 
We set a Hubbard $U$ parameter to 3.78 eV, which produced the lowest fixed-node energy in our previous DMC study for a single Pt atom adsorbed on the graphene surface~\cite{ahn21}. 
A plane-wave cutoff of 400 Ry and a $6 \times 6 \times 1$ Monkhorst-Pack grid~\cite{monkhorst76} were set to generate the PBE+$U$ orbitals. 
The optimization of Jastrow parameters was performed through variational Monte Carlo using the linear method proposed by Umrigar {\it et al.}~\cite{umrigar07},
followed by DMC calculations with a time step of $\tau = 0.005$ Ha$^{-1}$ where size-consistent T-moves were used for imaginary time projection to evaluate non-local pseudopotentials variationally~\cite{kim18}.
A norm-conserving scalar-relativistic pseudopotential for 18 valance electrons of $5s^{2}5p^{6}6s^{0}5d^{9}$ was employed for the Pt atom as in our previous DMC study~\cite{ahn21,ahn2023structural,ahn2024exploring,ahn2025stacking}, while the pseudopotentials by Burkatzki, Filippi and Dolg (BFD)~\cite{burkatzki07,burkatzki08} were used for other atomic species. 
When it comes to estimating density which does not commute with the Hamiltonian, an extrapolated estimator, $\rho_{\text{ext}} = 2\rho_{\text{DMC}}-\rho_{\text{VMC}}$~\cite{foulkes01}, was utilized in our study.

\begin{figure}
\includegraphics[width=4.0in]{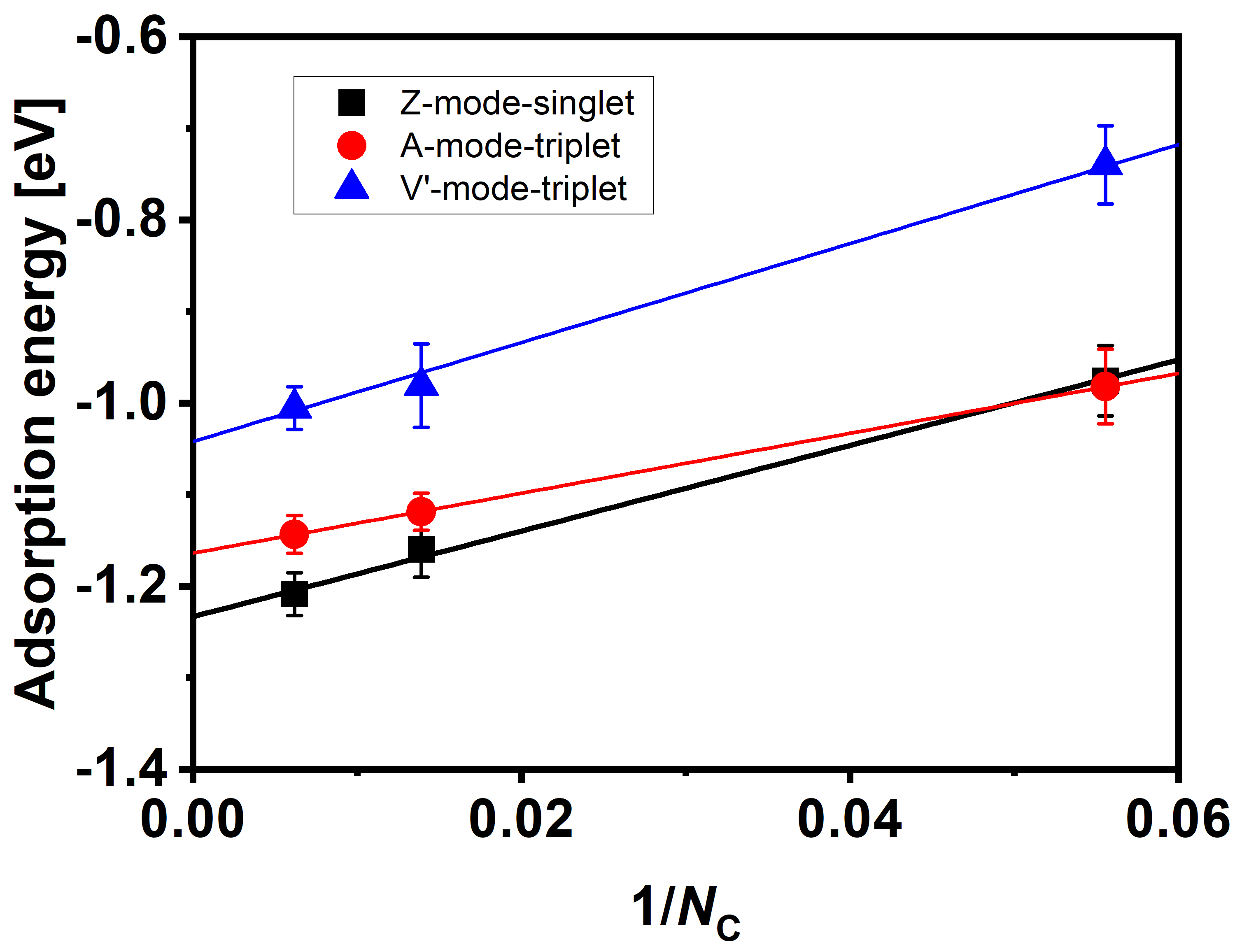}
\caption{(Color online) DMC adsorption energies of an O$_{2}$ molecule adsorbed on a graphene-supported Pt atom as functions of $1/N_{c}$ for three different configurations (Z-, A- and V$'$-modes), where $N_{c}$ represents the total number of carbon atoms in the supercell. The solid lines represent the extrapolations through linear regression fits.
 } 
\label{fig:O2_size_analysis}
\end{figure}

To minimize interactions between periodic images,
we configured a unit cell involving a single Pt atom adsorbed on a bridge site of a $3 \times 3 \times 1$ graphene cell plus a gas molecule attached to the Pt atom, resulting in a Pt–Pt separation of up to 7.38~\AA\ between adjacent periodic images.
%leading to a Pt–Pt separation of 7.38~\AA\ between periodic images.
%\hsadd{What is Pt-Pt distance on 3x3x1 cell and is it around the dissociation limit of Pt dimer? It looks better to put those values to claim the interaction is minimized.}
A vacuum distance along the $z$-axis perpendicular to the graphene surface was set to 20~\AA. One-body finite-size effects were minimized by applying twist-averaged boundary conditions~\cite{lin01} with 36 twisted angles for the smallest supercell.
We then conducted a standard $1/N$ extrapolation to mitigate higher-order finite-size effects. For this, our DMC calculations were performed for three different supercell sizes of $1 \times 1 \times 1$, $2 \times 2 \times 1$, and $3 \times 3 \times 1$.
Figure~\ref{fig:O2_size_analysis} shows the finite-size analysis of the O$_{2}$ adsorption energy though linear regression fits. 
%\hsadd{$N$ need to be defined. What is difference between $N$ and $N_{c}$? If you want to call the $1/N$ extrapolation as "standard", the reference will be needed.}

\section{Results and discussion}
\label{sec:results}

In this study, we investigate the adsorption of small gas molecules on a single Pt atom supported by pristine graphene. 
Accurate estimation of the molecular adsorption energies and geometries is essential for identifying the starting and final configurations in catalytic reactions that are directly concerned with the reaction paths and the energy barriers.
We used DMC calculations to examine the adsorption energetics of four gas molecules of CO, CO$_{2}$, O, and O$_{2}$, each of which is relevant to the CO oxidation process.
The molecular adsorption was studied for a Pt atom adsorbed on the bridge site, its most stable adsorption position on graphene. 
The DMC molecular adsorption energies were computed for several different configurations of the molecule-Pt-graphene complex, whose geometries had been optimized through PBE geometry relaxation calculations. 
Among DFT-optimized geometries using various density functionals, including LDA, PBE+D2, vdW-DF2 and rVV10, the PBE-optimized geometry turned out to yield the lowest DMC energy for a given configuration (see Fig. S1 and Table S1 in Supporting Information~\cite{SI} for more details), validating our use of the PBE-optimized geometries.
From this, we determined the most stable configuration as the one with the lowest DMC adsorption energy.
The molecular adsorption energy is defined by $E_{ad}=E_{\text{total}}-E_{\text{Pt-graphene}}-E_{\text{molecule}}$ where $E_{\text{total}}$, $E_{\text{Pt-graphene}}$, and $E_{\text{molecule}}$ represent the energies of the total complex, the Pt-graphene system, and
an isolated molecule, respectively.

\subsection{O$_2$ adsorption}
\label{subsec:O2}

\begin{figure}[t]
\includegraphics[width=6.5in]{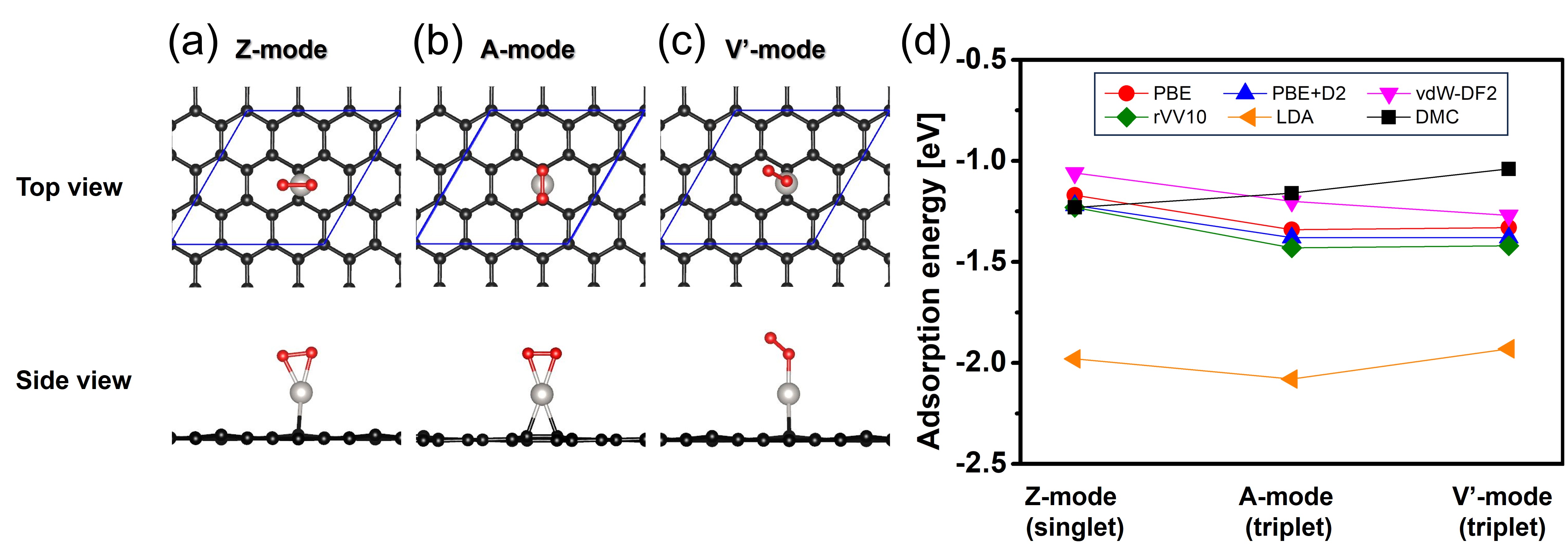}
\caption{(Color online) 
(a) to (c) Three stable configurations of an O$_{2}$-adsorbed Pt-graphene complex, where black, white, and red spheres represent carbon, platinum, and oxygen atoms, respectively.
Here the blue parallelograms indicate a $3 \times 3 \times 1$ graphene supercell. (d) presents DMC and DFT molecular adsorption energies for each configuration.} 
\label{fig:O2}
\end{figure}

In Fig.~\ref{fig:O2}, we present three selected configurations for an O$_{2}$ molecule adsorbed on a graphene-supported Pt atom, all of which were assessed to be stable in both PBE and DMC calculations (see Fig. S2 of Supporting Information~\cite{SI} for all the O$_{2}$ configurations considered).
Noting that an isolated O$_{2}$ molecule is in a triplet state, we considered both the singlet and triplet spin states, as well as the geometric degrees of freedom, for the entire O$_2$-Pt-graphene complex.
The configurations are labeled according to the orientation of the O$_2$ molecular axis relative to the graphene surface: the Z-mode corresponds to alignment along the zigzag direction, the A-mode along the armchair direction, and the V$'$-mode represents a configuration where the molecular axis is tilted with respect to the vertical direction of the graphene surface.
Both the Z- and A-modes are side-on configurations where the O$_{2}$ molecule form two chemical bonds with the Pt atom, while the V$'$-mode is an end-on configuration with only one oxygen atom participating in the O-Pt bond. In each configuration, the O-O bond length is observed to be elongated compared to the bond length of 1.208~\AA~in an isolated O$_{2}$ molecule, as detailed in Table S2 of Supporting Information~\cite{SI}.

As illustrated in Fig. S2 of the Supporting Information, our DMC calculations for the O$_2$ adsorption energy unveil that the singlet state is energetically preferred over the triplet state for the Z-mode, while the triplet state is more stable in the A- and V$'$-modes. 
The DMC adsorption energies of singlet Z-mode, triplet A-mode, and triplet V$'$-mode are estimated to be -1.23(2) eV, -1.16(1) eV, and -1.04(1) eV, respectively. From these results, we conclude that the singlet Z-mode is the most stable configuration for O$_2$ adsorption on the Pt-graphene complex.
The preference for different spin states in the side-on A- and Z-modes can be attributed to the different alignments of the O$_2$ molecular axis relative to the graphene sheet.  
In the Z-mode, the molecular axis is not parallel to the graphene sheet (see Fig.~\ref{fig:O2}), 
which breaks the orbital degeneracy and favors the paired spin configuration of the singlet state.
Conversely, in the A-mode, where the molecular axis is parallel to the substrate, the orbital hybridization between Pt and O atoms is symmetric.
This leads to degenerate highest-occupied levels, which favors the unpaired spin configuration of the triplet state.

We now compare our DMC results for O$_2$ adsorption on a graphene-supported single Pt atom with the corresponding DFT results, computed using several different exchange-correlation functionals, as shown in Fig.~\ref{fig:O2}(d) and Table S3 of Supporting Information~\cite{SI}. 
Both PBE and DMC calculations predict the same preferred spin state for each O$_2$ adsorption mode; however, PBE significantly overestimates the adsorption energies compared to DMC.
Unlike the DMC results, which identify the singlet Z-mode as the most stable configuration, 
our PBE calculations predict that the lowest-energy configuration is the triplet A-mode.
The previous spin-polarized PBE calculations of Tang {\it et al.}~\cite{tang12} also reported that the configuration parallel to the graphene surface, similar to our A-mode, was the most stable.
This discrepancy between DMC and PBE results highlights the limitation of PBE in accurately describing O$_2$ adsorption on a graphene-supported Pt atom, particularly due to its inability to capture many-body correlation effects which are critical in determining adsorption energetics of Pt-involved systems~\cite{ahn21,ahn2023structural,ahn2024exploring,ahn2025stacking}.
Given that the choice of initial reactant configurations can significantly influence reaction paths and activation barriers, our results suggest that more sophisticated computational methods beyond PBE are required to accurately determine the initial configurations of the CO oxidation processes on the Pt-graphene complex. 
Lastly, we observe that the inclusion of van der Waals (vdW) corrections does not provide much improvement over the PBE results, indicating that vdW interactions do not play a significant role in the adsorption energetics in this system.

%*****
\begin{figure}
\includegraphics[width=4.0in]{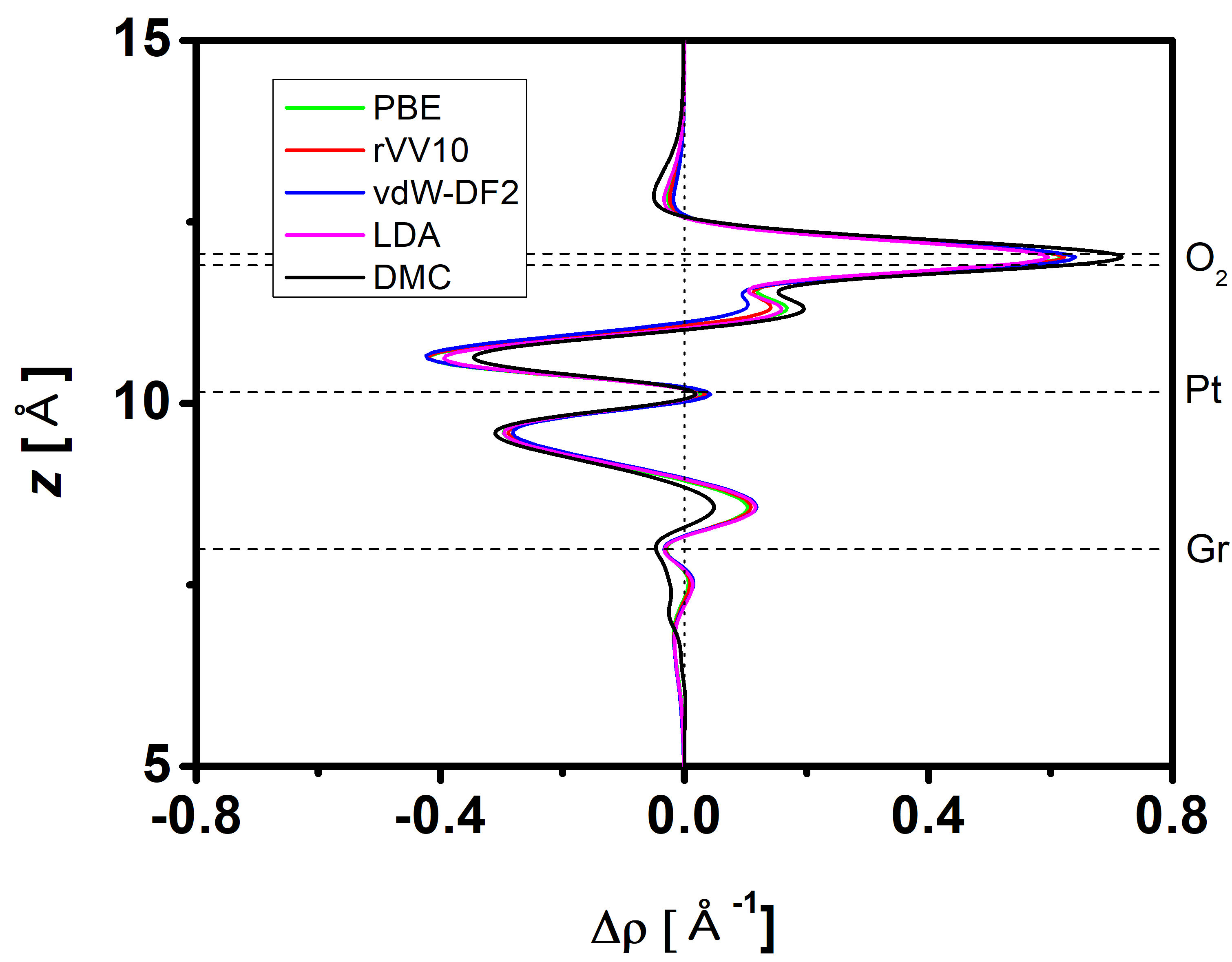}
\caption{(Color online) 
One-dimensional charge density differences, projected along the vertical $z$-axis, between the O$_2$-adsorbed Pt-graphene complex in the Z-mode and the Pt-graphene system plus an isolated O$_2$ molecule.
The horizontal dashed lines indicate the vertical positions of the graphene sheet, the Pt atom, and each oxygen atoms of the O$_2$ molecule, while the vertical dotted line corresponds to the zero charge density difference.
 } 
\label{fig:CDD}
\end{figure}

For further analysis of the O$_2$ adsorption. we now investigate charge density redistribution in the lowest-energy singlet Z-mode. 
Figure~\ref{fig:CDD} show DMC charge density difference, which were estimated using the extrapolated estimators (details provided in the Methodology section), along with the corresponding DFT results.
Here the charge density difference is defined by $\Delta \rho=\rho_{\text{Gr-Pt-O$_{2}$}}-\rho_{\text{Gr-Pt}}-\rho_{\text{O$_{2}$}}$ where $\rho$ represents the laterally-averaged one-dimensional charge density projected along the $z$-axis. Positive values indicate charge accumulation, while negative values correspond to charge depletion.
The DMC results show charge accumulation on the O$_{2}$ molecule, compensated by a migration of charge from the Pt-graphene complex, suggesting a charge transfer from the Pt-graphene complex to the O$_2$ molecule.
Comparing the DMC and DFT results reveals slight discrepancies, with DFT underestimating the amount of charge transfer from the Pt-graphene complex to the O$_2$ molecule. These discrepancies reflect the limitation of DFT as a mean-field method, contributing to the differences in adsorption energies between DFT and DMC.

%\ihadd{These species are distinguished by their O–O bond lengths: superoxo species typically have bond lengths of 1.25–1.35~\AA, while peroxo species range from 1.35–1.45~\AA. On the Pt(111) surface, the peroxo state is the most stable, similar to what is observed for DMC. Interestingly, Single-Atom Catalysts (SACs) can be considered analogs of coordination compounds, which implies that the hydrogen evolution reaction (HER) and oxygen evolution reaction (OER) can proceed through the formation of highly stable intermediates that are not usually observed on extended metal surfaces. A comprehensive understanding of these structural characteristics and their influence on reactivity is crucial for explaining catalytic behavior and for making accurate predictions about the performance of new catalysts.} %https://pubs.acs.org/doi/10.1021/acscatal.2c03020 
%https://link.springer.com/article/10.1007/s11244-023-01802-x
%https://www.sciencedirect.com/science/article/pii/S0920586123004339
%https://chemistry-europe.onlinelibrary.wiley.com/doi/10.1002/cphc.201000286

\subsection{Adsorptions of CO, CO$_2$, and Atomic Oxygen}
\label{sec:CO and CO2}

\begin{figure}[t]
\includegraphics[width=6.5in]{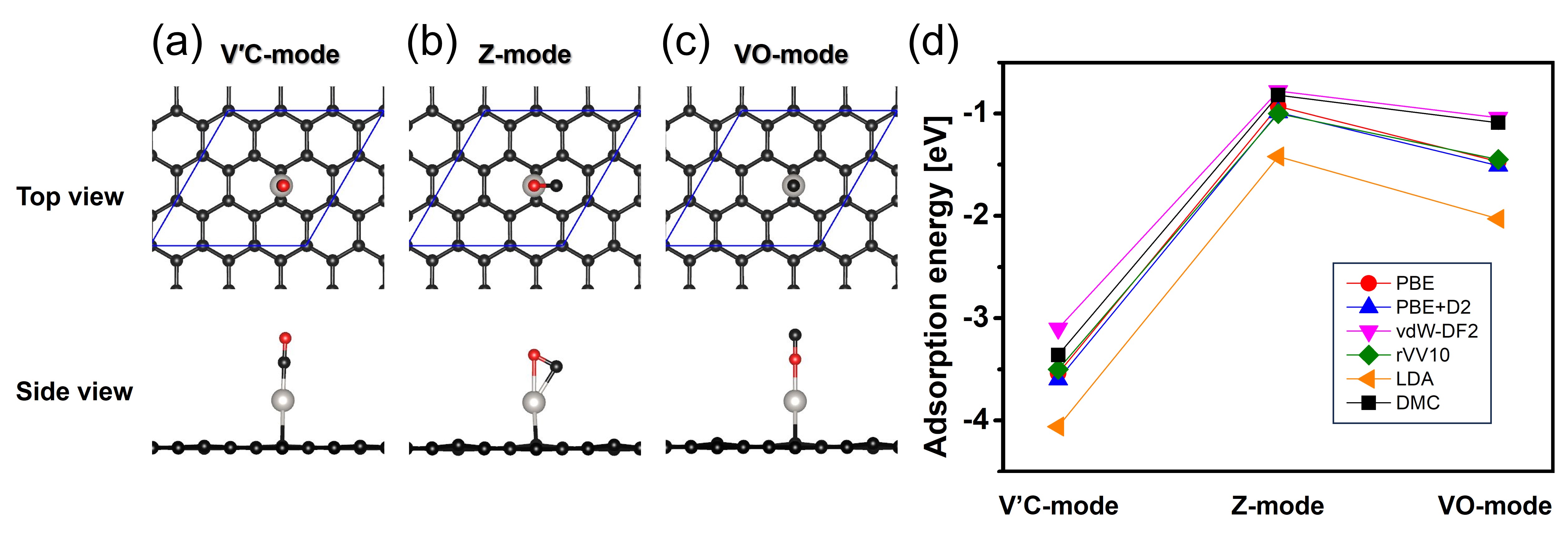}
\caption{(Color online) 
(a) to (c) Three PBE-optimized configurations ((a) to (c)) of a CO-adsorbed Pt-graphene complex, where black, white, and red spheres represent carbon, platinum, and oxygen atoms, respectively. Here the blue parallelogram indicates a $3 \times 3 \times 1$ graphene supercell. (d) presents DMC and DFT molecular adsorption energies for each configuration.
 } 
\label{fig:CO}
\end{figure}

For the CO adsorption on the Pt-graphene complex, as shown in Fig.~\ref{fig:CO}(a), the most stable configuration is found to be the V$'$C-mode, a tilted end-on configuration, where the CO molecular axis is tilted by 2$^{\circ}$ from the vertical direction. 
The DMC adsorption energy for CO in the V$'$C-mode is calculated to be -3.37(1) eV, which is slightly lower than the adsorption energy of -3.34(1) eV for the vertical end-on configuration without tilting. In contrast, the side-on Z-mode of Fig.~\ref{fig:CO}(b) has a much higher adsorption energy of -0.82(1) eV, while the VO-mode Fig.~\ref{fig:CO}(c), where the O atom of the CO molecule forms a bond with the Pt atom, has an adsorption energy of -1.09(2) eV.
These DMC results for CO adsorption energies across different adsorption modes, along with the corresponding DFT adsorption energies computed using various exchange-correlation functionals, are presented in Fig.~\ref{fig:CO}(d).
The lower stability of the VO-mode relative to the V$'$C-mode can be attributed to the spatial distributions of the highest-occupied CO molecular orbitals, which are concentrated on the carbon atom side. This concentration enhances the orbital overlap between the carbon and Pt atoms in the V$'$C-mode, leading to greater stability of that configuration.

\begin{figure}[t]
\includegraphics[width=6.5in]{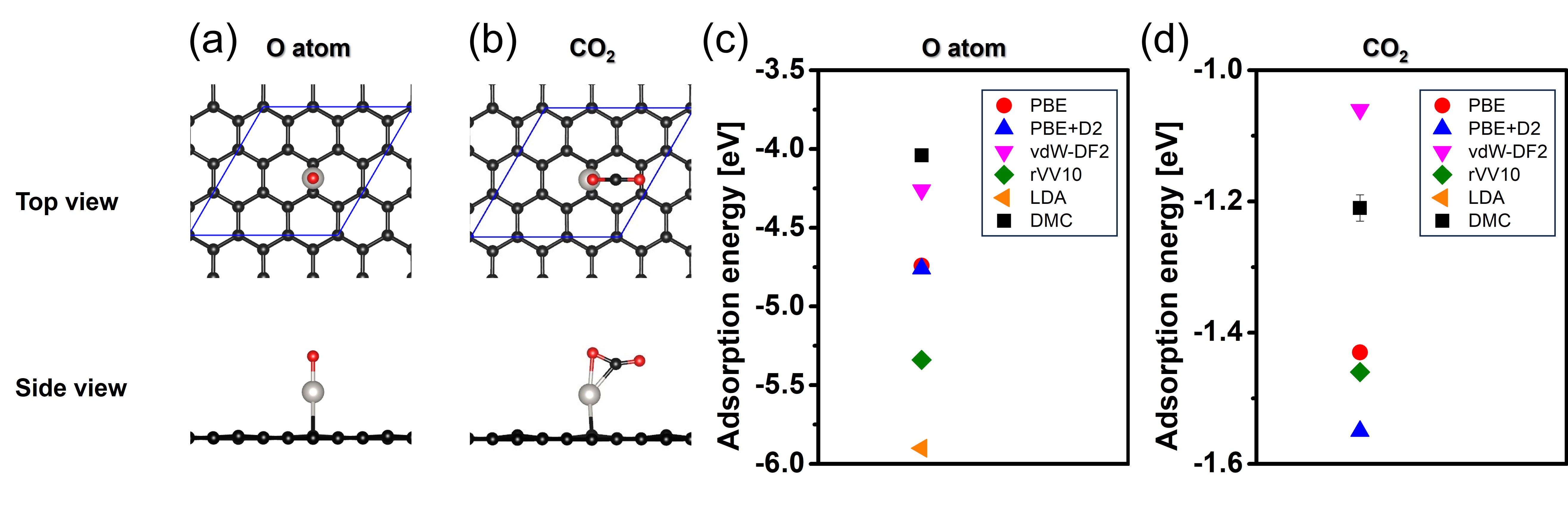}
\caption{(Color online) The most stable PBE-optimized configurations of (a) an O atom and (b) a CO$_{2}$ molecule adsorbed on the Pt-graphene complex, along with their respective adsorption energies in (c) and (d). Black, white, and red spheres represent carbon, platinum, and oxygen atoms, respectively. The blue parallelogram indicates a $3 \times 3 \times 1$ graphene supercell.
 } 
\label{fig:mol3}
\end{figure}

We now examine the adsorptions of an O atom and a CO$_2$ molecule on a graphene-supported single Pt atom, whose lowest-energy structures are shown in Fig.~\ref{fig:mol3}. 
The adsorption energetics of CO$_{2}$ are particulary important for understanding how product molecules are released from the Pt-graphene complex during CO oxidation reactions.
Three different PBE-optimized structures of a CO$_{2}$ adsorbed Pt-graphene complex are presented in Fig. S3 of Supporting Information. Among them, our DMC calculations identify a side-on configuration shown in Fig.~\ref{fig:mol3}(b) as the most stable, with the adsorption energy of -1.21(2) eV. In comparison, end-on configurations, whether tilted or aligned vertically, are found to be less stable. 

For atomic oxygen adsorption, the triplet state is found to be favored over the singlet state by 0.5(1) eV, with the DMC adsorption energy for the triplet state computed to be -4.04(1) eV (see Fig.~\ref{fig:mol3}(d)). 
%\hsadd{Is 0.5 eV from DMC calculation? If so, error bar needs to be added.}
From this, we estimate the bond dissociation energy of an O$_{2}$ molecule adsorbed on the Pt-graphene complex, defined as $E_{dissoc}=E_{\text{O$_{2}$-Pt-graphene}}-E_{\text{O-Pt-graphene}}-E_{\text{O}}$ where $E_{\text{O$_{2}$-Pt-graphene}}$, $E_{\text{O-Pt-graphene}}$, and $E_{\text{O}}$ are the total energies of the O$_{2}$-adsorbed Pt-graphene complex, the O-adsorbed Pt-graphene complex, and an isolated oxygen atom, respectively. The dissociation energy of O$_2$ in the singlet Z-mode is estimated to be 2.29(3) eV at the DMC level, which is significantly smaller than the corresponding DMC value of 5.08(1) eV for an isolated O$_{2}$ molecule. 
Notably, our DMC result for the isolated O$_{2}$ dissociation energy shows excellent agreement with the experimental value of 5.16 eV~\cite{afeefy2005neutral} which deviates considerably from the corresponding DFT predictions~\cite{perdew1996generalized,han2022evaluation}.

\subsection{Implications of DMC results for CO oxidation}

Previous DFT studies~\cite{tang12,liu14,tang15} have proposed the LH mechanism, involving the coadsorption of O$_2$ and CO molecules, as a plausible pathway for CO oxidation on Pt atoms supported by defective graphene. However, our study shows that the weaker binding of Pt atoms to pristine graphene, compared to defective graphene, inhibits the simultaneous adsorption of both CO and O$_2$ on a single Pt site. This implies that the LH mechanism is unlikely to occur on pristine graphene.

Moreover, our DMC calculations reveal a significant disparity between the adsorption energies of CO (-3.37(1) eV) and O$_{2}$ molecules (-1.23(2) eV), suggesting that the ER mechanism---where pre-chemisorbed O$_2$ molecules react directly with physisorbed CO---is also improbable. Instead, the results indicate that when a gas mixture of O$_{2}$ and CO molecules approaches the Pt-graphene surface, CO molecules preferentially adsorb onto the Pt sites, leading to CO poisoning. This strong CO adsorption can block the active sites, inhibit O$_2$ adsorption, and ultimately suppress the CO oxidation reaction on pristine graphene-supported Pt atoms.

On the other hand, the DMC adsorption energy of atomic oxygen (-4.04(1) eV) is significantly lower than that of CO. This suggests the potential formation of a stable O-Pt-graphene complex via O$_2$ dissociation or alternative pathways. Once formed, such a complex could facilitate the oxidation of CO through its reaction with a preadsorbed atomic oxygen, offering a plausible route to circumvent CO poisoning and enable catalytic activity.

\section{Conclusion}
Using DMC calculations, we have analyzed the adsorption energetics of gas molecules on a graphene-supported single Pt atom.
The results reveal significant discrepancies between DMC and DFT calculations in predicting the adsorption energies and molecular configurations, particularly for O$_2$ and CO molecules. Notably, DFT-PBE calculations often overestimate adsorption energies and fail to accurately predict the lowest-energy geometries and spin states. These discrepancies underscore the importance of considering many-body effects, which are crucial in accurately describing the adsorption energetics and determining the most stable configurations.

Our findings highlight the potential issues of CO poisoning, where CO preferentially adsorbs over O$_2$, potentially blocking active sites and hindering the efficiency of catalytic processes. This effect is a critical consideration for the design of Pt-based catalysts, especially in applications aimed at minimizing environmental impact through efficient CO oxidation. The detailed insights provided by our DMC calculations not only challenge the adequacy of conventional DFT methods for studying complex catalytic systems but also highlight the need for more sophisticated computational approaches that can accurately capture the intricate nature of many-body correlations. This work paves the way for the development of improved catalysts with enhanced activity and selectivity, offering significant potential for industrial applications in energy conversion and environmental remediation.

In conclusion, the integration of advanced many-body computational methods such as DMC into the study of catalytic materials is essential for pushing the boundaries of our understanding and for the design of the next generation of catalysts. The continued exploration of these methods will be vital in addressing the challenges of catalyst efficiency and in reducing the reliance on precious materials like platinum.

\section*{Data Availability}
\label{sec:conclude}

All codes, scripts, and data needed to reproduce the results in this paper can be found online at the Materials Data Facility [link to be added upon acceptance]. 

\section*{Acknowledgements}
\label{sec:acknowledgments}

%Y. Kwon was supported by the Basic Science Research Program (2018R1D1A1B07042443)through the National Research Foundation of Korea funded by the Ministry of Education.
%This paper was supported by Konkuk University in 2024.
Initial work by J.Ahn was supported by Konkuk University. J.Ahn also acknowledges support from U.S. Department of Energy, Office of Science, Office of Basic Energy Sciences, Computational Materials Sciences Award No. DE-SC0020177 for analysis and writing of the paper.
Calculations were performed by J. Ahn at Konkuk University.
H. Shin and A. Benali were supported by the U.S. Department of Energy, Office of Science, Basic Energy Sciences, Materials Sciences and Engineering Division, as part of the Computational Materials Sciences Program and Center for Predictive Simulation of Functional Materials.
An award of computer time was provided by the Innovative and Novel Computational Impact on Theory and Experiment (INCITE) program and was used to generate all DMC results. This research used resources of the Argonne Leadership Computing Facility, which is a DOE Office of Science User Facility supported under contract DE-AC02-06CH11357. We also acknowledge the support from the Supercomputing Center/Korea Institute of Science and Technology Information with supercomputing resources  (KSC-2023-CRE-0136) that were used for DFT calculations.

\bibliography{Pt_Gr}% Produces the bibliography via BibTeX.

\end{document}